\documentclass[apl,amsmath,amssymb,reprint]{revtex4-1}

\usepackage{graphics}

\begin{document}

	\preprint{AIP/123-QED}
	
	\title[Optical trapping and control of nanoparticles inside evacuated hollow core photonic crystal fibers]{Optical trapping and control of nanoparticles inside evacuated hollow core photonic crystal fibers}
	
	\author{David Grass}
	\email{david.grass@univie.ac.at}
	\affiliation{Vienna Center for Quantum Science and Technology (VCQ), Faculty of Physics, University of Vienna,
		A-1090 Vienna, Austria}
	\author{Julian Fesel}
	\affiliation{Vienna Center for Quantum Science and Technology (VCQ), Faculty of Physics, University of Vienna,
		A-1090 Vienna, Austria}
	\author{Sebastian G. Hofer}
	\affiliation{Vienna Center for Quantum Science and Technology (VCQ), Faculty of Physics, University of Vienna,
		A-1090 Vienna, Austria}
	\author{Nikolai Kiesel}
	\affiliation{Vienna Center for Quantum Science and Technology (VCQ), Faculty of Physics, University of Vienna,
		A-1090 Vienna, Austria}
	\author{Markus Aspelmeyer}
	\email{markus.aspelmeyer@univie.ac.at}
	\affiliation{Vienna Center for Quantum Science and Technology (VCQ), Faculty of Physics, University of Vienna,
		A-1090 Vienna, Austria}

	\date{\today}
	
	\begin{abstract}We demonstrate an optical conveyor belt for levitated nano-particles over several centimeters inside both air-filled and evacuated hollow-core photonic crystal fibers (HCPCF). Detection of the transmitted light field allows three-dimensional read-out of the particle center-of-mass motion. An additional laser enables axial radiation pressure based feedback cooling over the full fiber length. We show that the particle dynamics is a sensitive local probe for characterizing the optical intensity profile inside the fiber as well as the pressure distribution along the fiber axis. In contrast to previous indirect measurement methods we find a linear pressure dependence inside the HCPCF extending over three orders of magnitude from 0.2 mbar to 100 mbar. A targeted application is the controlled delivery of nano-particles from ambient pressure into medium vacuum.
	\end{abstract}
	
	\pacs{Valid PACS appear here}
	\maketitle
	Optically levitated nano-particles are a new paradigm in the development of high-quality mechanical resonators \cite{Yin2013}. This approach has been proposed as a route for ultra-sensitive force measurements \cite{Arvanitaki2013}, studies of stochastic out-of-equilibrium physics in the underdamped regime \cite{Lechner2013,Dechant2015}, and for room-temperature quantum optomechanics \cite{Chang2010,Romero-Isart2010a,Barker2010}. Early experiments by Ashkin have  demonstrated optical levitation of micrometer-scale dielectric objects in high vacuum \cite{Ashkin1976}. Recent efforts focus on sub-micron particles and have already demonstrated mechanical quality factors up to $10^8$ (see ref. \cite{Gieseler2013}), thereby surpassing those of state-of-the-art clamped nano-mechanical devices \cite{Imboden2014}. Further examples include the realization of zepto-Newton force sensing \cite{Ranjit2016}, a test of fluctuation theorems \cite{Gieseler2014a}, and optical feedback- and cavity-cooling \cite{Li2011,Gieseler2012b,Ranjit2015a,Kiesel2013b}. 
	
	Hollow-core photonic crystal fibers \cite{Cregan1999,Russell2006} (HCPCF) add a particularly intriguing instrument to the toolbox of levitated optomechanics. Free-space optical traps typically offer only small volumes of high light intensity, and in this respect limited capabilities for optical micro-manipulation. By contrast, HCPCF allow a tight transversal confinement of optical fields over the full distance of the fiber length, which can extend over several meters. Recently, precise control over micrometer sized particles in HCPCFs has been established in several experiments \cite{Benabid2002,Schmidt2012,Schmidt2013} and first sensing capabilities have been demonstrated \cite{Bykov2015}. Here we present new methods for optical micro-manipulation of particles inside HCPCF and extend the application regime to sub-micron sizes. Specifically, we demonstrate an optical conveyor belt for nano-particles inside a HCPCF, in which particles are optically trapped in a standing wave field and can be transported and precisely positioned along the fiber. We show three-dimensional (3d) read-out of the center-of-mass (COM) motion together with feedback control in axial direction. We demonstrate that a particle can be delivered into vacuum (0.2 mbar) with a pressure difference of three orders of magnitude between the fiber ends. We use the read-out and control capabilities to investigate the pressure distribution inside the HCPCF. 
	
	To establish the optical conveyor belt two counterpropagating laser fields with equal polarization and wavelength $\lambda_\mathrm{tr}$ excite the eigenmodes of a HCPCF, which are approximated with linear polarized modes \cite{Marcatili1964} ($\mathrm{LP}_\mathrm{ij}$).The resulting standing wave, mainly formed by the fundamental mode ($\mathrm{LP}_{01}$), creates an optical trap for Rayleigh particles with trapping potential $U=-\alpha I(r,z)/(2\varepsilon_0c)$ (particle polarizability $\alpha=4\pi\varepsilon_0a^2(\varepsilon-1)/(\varepsilon+2)$, particle radius $a\ll\lambda_\mathrm{tr}$, dielectric constant $\varepsilon$, vacuum permeability $\varepsilon_0$, speed of light $c$, intensity distribution $I(r,z)$, $r=\sqrt{x^2+y^2}$ and $z$ denote radial and axial direction inside the HCPCF, respectively). For a sufficiently deep potential the particle is trapped close to an intensity maximum through the gradient force \cite{Harada1996} $\vec F_\mathrm{grad}=-\nabla U$. By introducing a frequency detuning $\Delta\nu$ between the counterpropagating lasers the standing wave pattern moves along the fiber axis carrying the trapped particle at a velocity $v_z=\Delta\nu\lambda_\mathrm{tr}/2$, analogous to standing wave conveyor belt techniques in free space \cite{Kuhr2003,Cizmar2005}.
	
	For small displacements, the potential $U$ can be approximated as a 3d harmonic oscillator potential. The equation of motion for the trapped particle's COM motion along the fiber axis is
	\begin{align}
	\label{eq:equation_of_motion}
	\ddot z + \Gamma_p\dot z + \Omega^2 z =  \frac{F_\text{therm}}{m}
	\end{align}
	with $m$ the mass and $\Omega$ the mechanical frequency of the particle. Collisions with gas molecules result in a pressure dependent Stokes friction \cite{Beresnev1990,Li2011,Gieseler2012b} force $F=m\Gamma_p\dot z$ and in Brownian force noise $F_\text{therm}$. The latter is a zero-mean stochastic process $\langle F_\text{therm}\rangle=0$ satisfying the correlation function $\langle F_\text{therm}(t)F_\text{therm}(t')\rangle=\delta(t-t')2m\Gamma_pk_\mathrm B T_0$ for the case of a Markovian heat bath (Boltzmann's constant: $k_\mathrm B$,  environment temperature: $T_0=293\ $K).
	
	\begin{figure}[h!]
		\centering
		\includegraphics{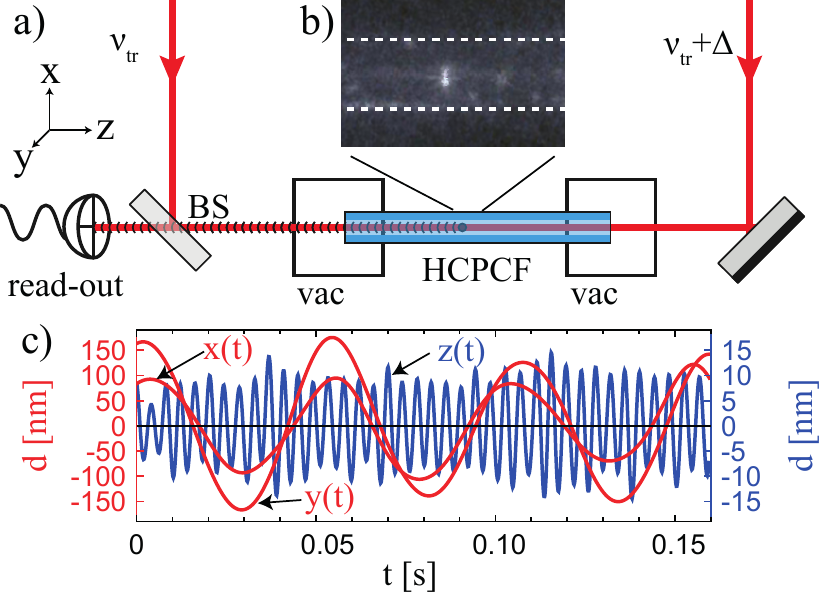}
		\caption{\label{fig:setup}a) Schematic drawing of the experimental setup: A hollow core photonic crystal fiber (HCPCF) connects two vacuum chambers (vac) whose pressures can be individually controlled. Two equally polarized, counterpropagating lasers with frequencies $\nu_\text{tr}$ and $\nu_\text{tr}+\Delta$ are focused into the HCPCF. For  $\Delta=0$ they form a standing wave optical trap for silica nano-particles. A beam splitter (BS) is used to split off 10\% of the clockwise propagating light for 3-dimensional detection of the particle motion (read-out). Particles from a nebulizer source (not shown) are trapped in front of the HCPCF; they can be transported along the fiber by detuning the clockwise propagating laser $\Delta\neq0$. b) Image of a $200\ $nm radius silica particle trapped inside the HCPCF (indicated with white dashed lines). c) 3d trajectory of a trapped nano-particle (z: along fiber axis, x,y: radial directions)}
	\end{figure}
	
	In our experiment (figure \ref{fig:setup}a), two vacuum chambers are connected through a 15 cm long HCPCF (HC-1060, NKT Photonics). Each chamber has an independent pressure control and one chamber is connected to a nebulizer that supplies airborne silica nano-particles from an isopropanol solution (Appendix \ref{ch:suppA}). The counterpropagating fields for trapping and transport are derived from a single-frequency Nd:YAG laser ($\lambda_\mathrm{tr}=1064\ $nm). Each of the beams is individually frequency shifted by an acousto-optical modulator (AOM), which allows control over their relative detuning $\Delta\nu$. The AOMs are driven by a dual-frequency source where both channels are synthesized from the same master oscillator allowing a relative detuning $\Delta\nu$ between the outputs in discrete steps of 10 kHz. This allows to realize step-sizes for the particle motion of approx. $12\ \mu$m. Fine-grained positioning on the sub-micron scale (around $0.5\ \mathrm\mu$m) is achieved using a manual position stage to change the optical path length difference by $\Delta l$, thereby displacing the standing wave by $\Delta l/2$. 
	
\begin{figure}[h!]
		\centering
		\includegraphics{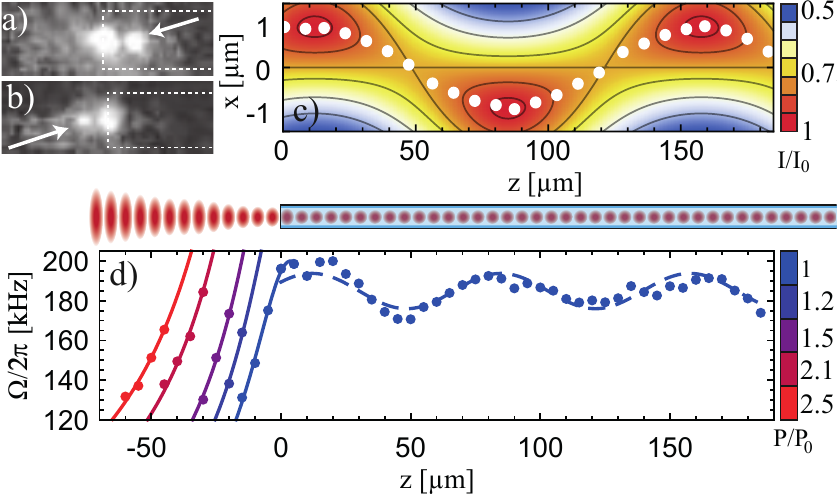}
		\caption{\label{fig:getting_it_out} Image of a $200\ $nm radius silica particle (white arrow) trapped inside a) and outside b) the HCPCF (indicated by the white dashed line). Note that additional scattering occurs at the fiber cleave. c) Simulated intensity distribution inside the HCPCF due to interference between the fundamental $\mathrm{LP}_{01}$ and a higher order $\mathrm{LP}_{11}$ mode. Different trapping positions (white dots) also exhibit different axial frequencies. d) Axial mechanical frequency of a trapped particle inside ($z>0$) and in front of the fiber ($z<0$). Inside the fiber, the modulation of the mechanical frequency  $\Omega$ changes with the distance of the trap position from the axis as shown in c). Outside the fiber the axial frequency decreases with the distance from the fiber tip due to the expansion of the mode. Increasing the laser power results in extended trapping distances. Here, the particle is stably trapped up to $65\ \mathrm\mu$m in front of the HCPCF at a pressure of $p=0.3\ $mbar.}
	\end{figure}

	A particle trapped inside the HCPCF scatters light partially into the eigenmodes and partially through the fiber walls. This allows 3d-monitoring of the COM position relative to the standing wave trap maximum, and also of the absolute position in the fiber. Figure \ref{fig:getting_it_out} shows an example of the light observed on a laterally mounted CCD-camera when the particle is trapped inside (figure \ref{fig:getting_it_out}a) and outside (figure \ref{fig:getting_it_out}b) the fiber. The interference between the trapping laser and the light scattered into the HCPCF eigenmodes is used to detect the COM motion of the particle in all three directions. To this end, 10\% of the transmitted light of one fiber end is used for the read-out (figure \ref{fig:setup}a). Light scattered into the fundamental mode due to the axial motion of the particle results in an intensity modulation of the detected field. The radial motion causes light scattering into higher order fiber modes breaking the radial symmetry of the field distribution. They can therefore be observed on a quadrant-diode-like detection configuration (Appendix \ref{ch:suppB}). Figure \ref{fig:setup}c shows a 150 ms section of a 3-dimensional displacement time trace of a trapped nano-particle.
	
	The geometry of the optical trap has a periodicity $\lambda_\mathrm{sw}=2\pi/(2\beta_{01})\approx 0.53\ \mathrm\mu$m that is determined by the wave-vector $\beta_{01}$ of the fundamental mode. A weak excitation of the higher-order $\mathrm{LP}_{11}$ mode with a different wave vector $\beta_{11}=\beta_{01}-\Delta\beta$ results in a modulation of the radial intensity profile of the optical trap. Specifically, due to interference between the eigenmodes, the radial position of the intensity maxima oscillate along the z-axis with a beat-note modulation of $\eta_\mathrm{mod}=2\pi/\Delta\beta$. This is illustrated in figure \ref{fig:getting_it_out}c, where the white dotted line marks the intensity maximum, along which particles would be trapped\footnote{The standing wave contribution is not shown as it modulates the pattern on a much shorter length scale}. Note that the same interference effect gives also rise to a modulation of the axial intensity profile with a periodicity of $\eta_\mathrm{mod}/2$.
	
	A measurement of the mechanical frequency distribution along the fiber axis confirms this multi-mode behavior (figure \ref{fig:getting_it_out}d). For trap positions inside the HCPCF a sinusoidal fit (blue dashed line) to the position dependent frequency yields a modulation with a period of $(72.3\pm1.6)\ \mathrm\mu$m. This corresponds to a core radius of $r_\text{co}=(4.2\pm.01)\ \mathrm\mu$m of the fiber, which is in agreement with the manufacturer specification $r_\text{co}=(5\pm1)\ \mathrm\mu$m. Furthermore, the amplitude of the modulation allows to estimate the power ratio between the higher order mode ($\mathrm{LP}_{11}$) and the fundamental mode ($\mathrm{LP}_{01}$) with $\mathrm{P_{11}/P_{01}}=(0.10\pm0.02)$. When the particle is moved out of the fiber the intensity decreases due to the increase in beam diameter. As a consequence, both the trap frequency and trap confinement decrease with increasing distance of the particle from the HCPCF exit face. We compensate this effect by increasing the power of the trapping laser while moving the particle out. The corresponding mechanical frequencies are shown in figure \ref{fig:getting_it_out}d ($z<0$) with different colors for different trapping beam powers and fits to Gaussian beam envelopes (solid lines). The waist of the mode is a free fit parameter and is determined as $w_0=(2.31\pm0.05)\ \mu$m. This value is smaller than the waist of the fundamental fiber mode ($3.15\ \mu$m) which we attribute to a tighter focus of the beam that is coupled into the HCPCF. All shown data were obtained at a pressure of $p=0.2\ $mbar, where the particle can still be stably trapped at a distance of $65\ \mu$m in front of the fiber. At lower pressures the particle is lost from the trap, both inside and in front of the HCPCF. This is a common phenomenon \cite{Kiesel2013b,Monteiro2013,Tongcang2011,Moore2014a,Ranjit2015a,Ranjit2016} that is poorly understood so far and likely related to residual noise in the trapping potential. For the specific purpose of interfacing the particle with another optical field, like a cavity mode \cite{Mestres2015}, the distance from the fiber is an important benchmark to avoid scattering or shadowing effects. At higher pressures ($p\approx 5$mbar) we were able to trap particles at distances up to  $160\ \mathrm\mu$m in front of the HCPCF (Appendix \ref{ch:suppC}).
	
	\begin{figure}[h!]
		\centering
		\includegraphics{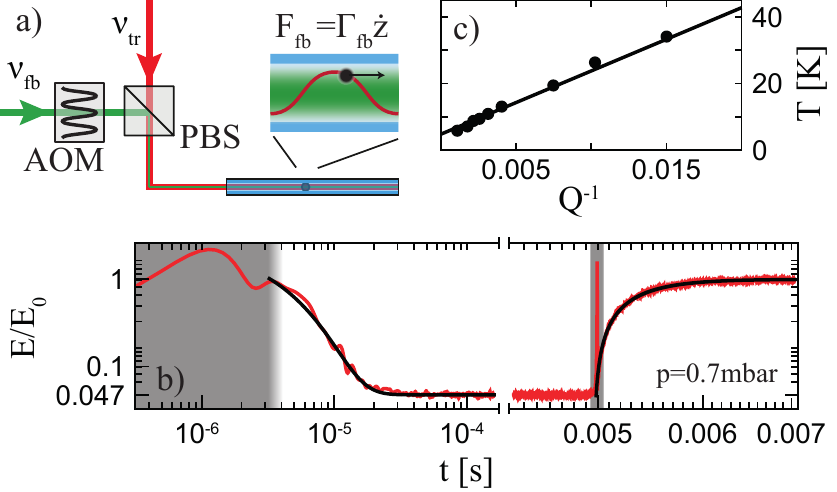}
		\caption{\label{fig:feedback}Feedback cooling and rethermalization of the nano-particle COM motion inside a HCPCF: a) A laser for feedback cooling ($\nu_\mathrm{fb}$) is superimposed on a polarizing beam splitter (PBS) with the trapping laser. Its intensity is modulated with an acusto-optical modulator (AOM) to cool the COM motion of a trapped particle. b) Time dependence of the potential energy (red curve) of the nano-particle COM motion when feedback cooling is switched on (left half) and off (right half). Fitting an exponential dependence to the transient processes (black curves) determines the values for damping and cooling rates $\Gamma_p$ and $\Gamma_\mathrm{fb}$. The two peaks (within the grey shaded areas) are caused by the switching processes. c) Relation between pressure dependent mechanical quality and effective mode temperature for constant feedback strength.}
	\end{figure}
	
	Next, we determine the damping rate $\Gamma_p$ of the particle COM motion. Depending on the pressure regime under investigation we either perform energy relaxation in the time domain or linewidth measurements in the frequency domain (Appendix \ref{ch:suppD}). While the latter method is limited by power drifts in the trap laser (and hence drifts in the trap frequency) to a regime where the linewidth $\Gamma_p$ is above 4 kHz, the first method requires feedback cooling and therefore works well for underdamped motion (i.e., low pressures). We realize feedback cooling with an additional laser ($\lambda_\mathrm{fb}=1064\ $nm) that is coupled into the HCPCF (figure \ref{fig:feedback}a). To avoid interference effects with the trapping laser it is well separated in frequency ($\nu_\text{fb}\neq\nu_\text{tr}$) and orthogonally polarized. Modulating the feedback laser power proportional to the velocity of the oscillator around a constant offset generates a feedback force $F_\mathrm{fb}= m\Gamma_\text{fb} \dot z$ on the particle at a new equilibrium position (Appendix \ref{ch:suppE}). Such a force modifies the friction term to $m(\Gamma_p+\Gamma_\text{fb})\dot z$ and hence, can be used for cooling of the COM motion, as already demonstrated in \cite{Ashkin1977,Cohadon1999}. When applied in 3d, feedback cooling is known to allow trapping in the high vacuum regime \cite{Li2011,Gieseler2012b}.
	
	Without any feedback control the COM motion of the particle is in thermal equilibrium and has the mean potential energy $E_0=1/2k_\mathrm B T_0=1/2m \Omega^2\langle z^2\rangle$. When the feedback is switched on, energy is extracted from the COM motion decaying as $E(t)=E_c+(E_0-E_c)e^{-\Gamma_\text{fb}t}$ to a lower value $E_c<E_0$ (figure \ref{fig:feedback}c). After switching off the feedback cooling, the system relaxes back to thermal equilibrium. During this transient process the time evolution can be approximated by an exponential \cite{Pinard2001} (Appendix \ref{ch:suppF}) yielding $E(t)=E_0-(E_0-E_c)e^{-\Gamma_p t}$. For each data point we calculate the potential energy as ensemble average over approximately 2000 switching processes. Figure \ref{fig:feedback}b shows an explicit example for the energy curve at $p=0.7\ $mbar. The relevant parameters $\Gamma_p$, $\Gamma_\mathrm{fb}$, $E_0$ and $E_c$ are obtained by fitting the exponential models.
	
	The corresponding effective temperature attained during the feedback process is $T_\mathrm{eff}=T_0E_c/E_0=T_0 \Gamma_p/(\Gamma_p+\Gamma_\mathrm{fb})+T_\mathrm{ro}$ where $T_\mathrm{ro}$ is a residual offset caused by noise in the position detection \cite{Poggio2007}. In our current implementation, we find $T_\mathrm{ro}=(4.83\pm 0.28)\ $K, constrained mainly by the shot-noise limited read-out and hence by the signal-to-noise ratio. The minimal effective temperature is $T_\mathrm{eff}=(5.87\pm 2.12)\ $K. Further improvement of the signal-to-noise ratio should enable feedback cooling to much lower temperatures \cite{Vovrosh2016,Jain2016}, and eventually, in combination with optimal filtering, into the quantum ground state of motion \cite{Wieczorek2015,Genes2008,Rodenburg2016}.
	\begin{figure}[h!]
		\centering
		\includegraphics{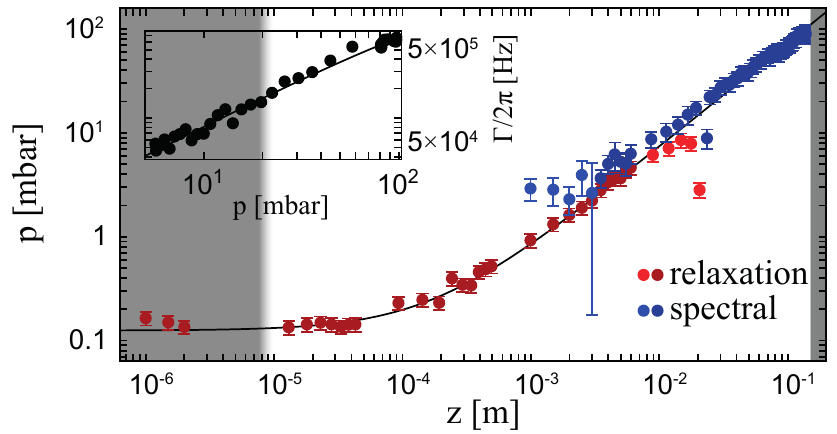}
		\caption{\label{fig:pressure}Pressure distribution measured along the HCPCF. A particle is trapped in front of the HCPCF at a pressure of $p\approx0.2\ $mbar (grey shaded area left) while the second vacuum chamber is at $p=10^2\ $mbar (grey shaded area right). The particle is moved in steps towards the second chamber. At each position the damping $\Gamma_p$ is measured via the energy relaxation method (red data points) or via a spectral evaluation (blue data points). The black curve is a linear fit to the data. Inset: Pressure calibration. The particle is trapped in front of the HCPC where local pressure and particle linewidth are independently measured by a vacuum gauge and optical read-out, respectively. The relation between damping $\Gamma_p$ and pressure $p$ is measured (points) and fitted (solid line).}
	\end{figure}
	
	The ability to measure the local damping rate of the particle motion enables us to use it as a nano-scale pressure sensor. It turns out that directly determining the pressure gradient inside a HCPCF that interconnects two reservoirs at different pressure, is a non-trivial task. It is related to the more general question of how pressure is distributed over a range that covers different flow regimes \cite{Sharipov1998}. In our case, the HCPCF establishes a 15 cm long channel of $10\ \mu$m diameter between two different pressure reservoirs: one at $p=0.2\ $mbar ($Kn\approx 30$) and the second at $p=10^2\ $mbar ($Kn\approx 0.06$). Here $Kn=\lambda_\mathrm{free}/r_\mathrm{co}$ is the Knudsen number, which characterizes the flow regime by comparing the mean free path of air molecules $\lambda_\mathrm{free}$ with the channel radius $r_\mathrm{co}$. To relate the mechanical damping $\Gamma_p$ of the particle COM motion to the local pressure $p$, we calibrate a nano-particle trapped at the edge of the HCPCF inside one vacuum chamber with the reading of a standard vacuum gauge that is connected to this chamber. The calibration data is shown in the inset of figure \ref{fig:pressure}. For the actual measurement of the pressure distribution we launch the particle approximately $15\ \mu$m in front of the HCPCF at the low-pressure end ($Q\approx 1400\pm 10$) and shuttle it towards the second chamber, i.e., the high-pressure end. The red data points shown in figure \ref{fig:pressure} correspond to energy relaxation measurements and the blue data points to a spectral measurement of the oscillator linewidth, as described above. The overlap region shows the consistency of the two measurement methods. The solid black line is a linear fit to the data excluding those points where the relaxation method (light red) and those where the spectral evaluation (light blue) breaks down.
	
	The data is consistent with a linear pressure distribution. For the simple case of molecular flow ($Kn>10$), this is expected, essentially because the mass transport is a pressure-independent diffusion process (Appendix \ref{ch:suppG}). 
	Interestingly, we find that the linear dependence even extends beyond the pressure of approximately 60 mbar, where we already enter the slip flow regime ($Kn<0.1$). This is in contrast to previous experimental studies of mass flow for a larger pressure difference in an otherwise comparable environment where a non-linear pressure dependence has been derived \cite{Yang2009}. Our result clearly contradicts this expectation and will be subject to further investigation.
	
	In conclusion, we have demonstrated optical trapping, position read-out and transport of particles with $200\ $nm radius over a distance of 15~cm inside a hollow-core photonic crystal fiber. Using an optical conveyor belt we show micron-scale positioning inside and in front of the fiber. We determine the pressure distribution inside the HCPCF with a core radius of $4.2\ \mathrm\mu$m for a pressure gradient of approximately $10^4\ $mbar/m between the fiber ends. In contrast to previous indirect predictions we find a linear pressure dependence in the molecular flow regime that even extends into the slip flow regime. The evaluation is based on spectral analysis for high pressures and equilibration from a feedback-cooled steady state for low pressures. The excellent level of control achieved in our experiment shows that HCPCFs provide a versatile tool for optical micro-manipulation of nano-particles.
	
	As a first relevant application of this method we envision delivery of nano-particles into a high-vacuum environment, well controlled in arrival time, position and effective temperature. This may enable further applications for force sensing, cavity optomechanics or matter-wave interferometry \cite{Bateman2013b,Haslinger2013}
	
	\begin{acknowledgments}
		We would like to thank U. Delic, O. A. Schmidt, T. G. Euser and P. Russel for stimulating discussions. We acknowledge funding from the European Commission via the Collaborative Project TherMiQ (Grant Agreement 618074) and the ITN cQOM, from the European Research Council (ERC CoG QLev4G) and from the Austrian Science Fund FWF under projects F40 (SFB FOQUS). D.G. is supported by the FWF under project W1210 (CoQuS).
	\end{acknowledgments}

		%
		%

		\appendix
		\section{Particle preparation and loading}
		\label{ch:suppA}
		The nano-particles used in this experiment are $\mathrm{SiO_2}$-F-0.4 from microParticles GmbH with a diameter $d=387\ \mathrm{nm}$. They come in a water solution with a 10\% mass concentration. The particles are diluted with Isopropanol to a mass concentration of $10^{-7}$. An Omron Air U22 asthma spray \cite{Burnham2006} is used in a nitrogen environment to nebulize the Isopropanol-particle solution. A cloud of airborne nano-particles from the nebulzer is sucked into the vacuum chamber. When a particle is trapped in front of the HCPCF the optical conveyor belt is switched on and particles are transported into the HCPCF.

		\section{Read-out of the particle motion}
		\label{ch:suppB}
		The readout of the particle COM motion inside the HCPCF is based on interference between scattered light from the particle and the fundamental $\mathrm{LP}_{01}$ trapping mode. We treat the nano-particle as dipole scatterer $E_\textrm{dp}$  which is preeminently excited by the fundamental $\mathrm{LP}_{01}$ trapping mode. Here, the axial readout ($z$) relies on scattering into the $\mathrm{LP}_{01}$ mode, and the transversal readout ($x,y$) on scattering into the $\mathrm{LP}_{11}$ mode, respectively. 
		\begin{figure}[h!]
			\centering
			\includegraphics{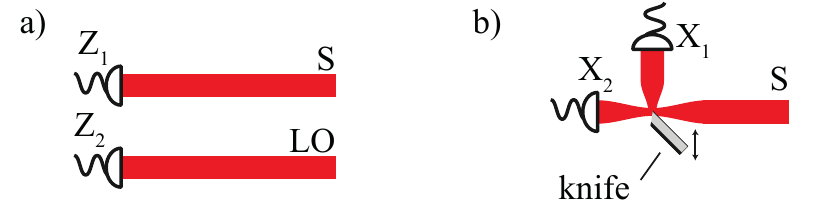}
			\caption{\label{fig:supp_read_out}a) Light that interacted with the nano-particle ($S$), sampled from the BS, is detected with photodiode $Z_1$ and a local oscillator ($LO$) is detected with photodiode $Z_2$. The difference between both photocurrents is proportional to the axial particle displacement. b) The signal ($S$) is focused on a knife edge such that 50\% of $S$ is reflected and send on photodiode $X_1$ and 50\% is transmitted on photodiode $X_2$. The knife position is locked to minimize $|X_1-X_2|$ and the difference photocurrent between both diode is proportional to the radial displacement of the nano-particle.}
		\end{figure}
		
		Let us assume the particle is axially displaced by $\delta z$. The mode overlap between scattered light from the particle and the fundamental mode is $\eta_{01}(\delta z)=\langle\mathrm{LP}_{01},E_\mathrm{dp}\rangle=|\eta_{01}|e^{i\beta_{01}2\delta z+i\phi_{01}}$ with $|\eta_{01}|,\phi_{01}$ independent of $\delta z$. The superposition of both fields arriving at detector $\mathrm{Z_1}$ is 
		\begin{align}
		\label{eq:axial_read_out}
		I_\mathrm{Z1}	&=			\frac{\varepsilon_0 c}{2}\big|\mathrm{LP}_{01}+\eta_{01}\mathrm{LP}_{01}\big|^2 							\nonumber 	\\
		&\approx	I_0-2I_0|\eta_{01}|(\beta_{01}-k)\delta z			 
		\end{align}
		with $I_0$ the intensity in the trapping mode and for $\phi_{01}-\pi/2,|\eta_{01}|^2\ll1$. The first term of equation \ref{eq:axial_read_out} is constant and the second is a signal proportional to the axial displacement $\delta z$ of the nano-particle. The detector $\mathrm{Z_1}$ integrates the intensity given in equation \ref{eq:axial_read_out}. As shown in figure \ref{fig:supp_read_out}a, a fraction of the trapping laser (LO) that does not interact with the nano-particle with the intensity distribution $I_0$ is integrated by a second detector $\mathrm{Z_2}$ such that the difference signal between both detectors results in the signal
		\begin{align}
		S_\mathrm{Z}	&=	S_\mathrm{Z_2}-S_\mathrm{Z_1} \\
		&=	2P_0|\eta_{01}|(\beta_{01}-k)\delta z
		\end{align}
		proportional to the particle displacement $\delta z$.
		
		The radial read-out relies on interference between scattered light from the particle into the higher order $\mathrm{LP}_{11}$ mode and the trapping mode. A particle moving radially (without loss of generality we assume motion along the $x$ axis, the treatment in the orthogonal direction y is analogous) excites the antisymmetric $\mathrm{LP}_{11}$ mode with a position dependent phase and amplitude $\eta_{11}(\delta x)=\langle\mathrm{LP}_{11},E_\mathrm{dp}\rangle=E_{01}(a+ib)\delta x$. The interference between the excited mode by the particle and the fundamental mode is
		\begin{align}
		I_\mathrm{X}	&=			\frac{\varepsilon_0c}{2}|\mathrm{LP}_{01}+\eta_{11}\mathrm{LP}_{11}|^2 	\\
		&=			I_0 + |\eta_{11}|^2\varepsilon_{01}^2 + \varepsilon_0 c E_{01}\Re\{\eta_{11}^*e^{i\Delta\beta z}\} \\
		&\approx	I_0 + \varepsilon_{01}\varepsilon_{11}\varepsilon_0 c E_{01}^2|\eta_{11}|\sin(\Delta\beta z+\phi_{11})\delta z
		\end{align}
		with $|\eta_{11}|=\sqrt{a^2+b^2}$ and $\phi_{11}=\arctan(b/a)$. It is important to note that the first term $I_0=\varepsilon_0 c/2E_0^2\varepsilon_{01}^2(x,y)$ is symmetric with respect to the x-axis and the second term $\propto\varepsilon_{01}(x,y)\varepsilon_{11}(x,y)$  is anti-symmetric with respect to the x-axis. Here $\varepsilon_{01}(x,y)$ and $\varepsilon_{11}(x,y)$ are the radial field distributions of the $\mathrm{LP}_{01}$ and $\mathrm{LP}_{11}$ mode. Secondly the term proportional to the displacement $\delta x$ also depends on the axial particle position $z$. Therefore, the sensitivity of the radial read-out is modulated with the wavelength of $2\pi/\Delta\beta\approx208\ \mu$m. 
		
		As shown in figure \ref{fig:supp_read_out}b the intensity distribution $I_\mathrm{X}$ is split on a knife-edge such that half of the mode is reflected to detector $\mathrm{X_1}$ and half of the mode is transmitted to detector $\mathrm{X_2}$. The difference signal between both detectors is 
		\begin{align}
		S_\mathrm{X}	&=			S_\mathrm{X1} - S_\mathrm{X2} =	\int\limits_{-x_0}^0dA\ I_\mathrm{X} - \int\limits_0^{x_0}dA\ I_\mathrm{X}\\
		&\propto	P_0|\eta_{11}|\sin(\Delta\beta z+\phi_{11})\delta z
		\end{align}
		with $x_0$ the extend of the detector. Altogether, the symmetric part of $I_\mathrm{X}$ vanishes and the anti-symmetric part stays due to the integration and results in a signal direct proportional to the radial particle motion $\delta x$. The read-out along the $y$-direction has the same underlying principle and, accordingly, a second knife-edge that is rotated by $90^\circ$.
		
		In order to monitor all three directions simultaneously, the light sampled at the beam splitter (BS, figure \ref{fig:setup}a, main text) is split into three parts for each spatial direction such that all can be monitored simultaneously, indicated with $S$ in figure \ref{fig:supp_read_out}.
		
		\section{Trapping in front of HCPCF at high pressure}
		\label{ch:suppC}
		\begin{figure}[h!]
			\centering
			\includegraphics{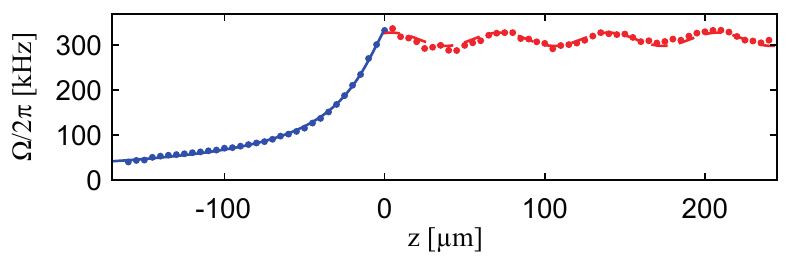}
			\caption{\label{fig:supp_getout}The mechanical frequency of a trapped nano-particle at $p\approx 5\ $mbar is recorded for different positions inside the HCPCF (red data points) and outside the fiber (blue data points). The modulation inside resembles the modulation of the intensity inside the fiber. The decrease of the mechanical frequency is caused by the diverging laser beam outside of the HCPCF. The solid lines are fits to the data.} 
		\end{figure}
		As mentioned in the main text the largest distance between HCPCF end and the position a particle can be stably trapped in front of the HCPCF is an important benchmark of our experiment. The data shown in the main text was taken at the lowest pressure at which we could keep a particle trapped. At higher pressures ($p\approx 5\ $mbar) we can increase the distance where a particle can be trapped to $160\ \mu$m from the HCPCF tip (Figure \ref{fig:supp_getout}).

		\section{Linewidth determination in the frequency domain}
		\label{ch:suppD}
		The spectrum of damped, thermally driven harmonic oscillator is \cite{Aspelmeyer2014a}
		\begin{align}
		\label{eq:spectrum}
		\begin{split}
		S_\mathrm x(\omega)	&=	\frac{2k_\textrm B T \Gamma_p}{\pi m}\frac{1}{(\Omega^2-\omega^2)^2+(\Gamma_p+\Gamma_\mathrm{fb})^2\omega^2}.		
		\end{split}
		\end{align}
		Equation \ref{eq:spectrum} has a peak at the mechanical frequency $\Omega$ with a linewidth $\Gamma=\Gamma_p+\Gamma_\mathrm{fb}$. Both values can be obtained from fitting. To resolve the linewidth $\Gamma$ from the spectrum a measurement time of at least $t\gtrsim1/\Gamma$ is necessary. However,in our experiment the mechanical frequency fluctuates due to drifts in the trapping laser, leading to inhomogeneous broadening of the spectrum. This effect becomes significant for a linewidth $\Gamma\lesssim4\ $kHz which we reach at $p\approx 4$ mbar. Below that pressure we switch to energy relaxation measurements.

		\section{Feedback control and center of mass motion temperature}
		\label{ch:suppE}
		The scattering force \cite{Harada1996} of the feedback laser is $F_\mathrm{scatt}=\frac{\sigma_\text{scatt}}{c}I(r,z)$ with $\sigma_\mathrm{scatt}=\frac{8\pi k^4a^6}{3}(\varepsilon-1)^2/(\varepsilon+2)^2$ the scattering cross section, $k=2\pi/\lambda$ the wave vector, $\varepsilon$ the dielectric constant of the particle, $c$ the speed of light and $I(r,z)$ the intensity distribution of the electromagnetic field.
		
		The scattering force pushes a trapped particle away from an intensity maximum of the standing wave. For a constant power $P_\text{fb}=P_0$ a trapped particle finds a new equilibrium position $z_0$ where $F_\text{scatt}(z_0)=F_\text{grad}(z_0)$. Modulation of the feedback laser around $P_0$ either pushes the particle further away from $z_0$ or the gradient force pulls the particle back towards intensity maximum of the standing wave. By modulating the laser power we can therefore effectively apply a force in both directions for a particle trapped at the new equilibrium position. Modulating the feedback laser proportional to the particle velocity results in a force $F_\text{fb}= m\Gamma_\text{fb} \dot z$.
		
		The read-out of the axial COM motion is used for feedback control. The voltage signal from the detector is highpass filtered (corner frequency $\nu_\mathrm{hp}\approx 50\ $kHz) and connected to a feedback circuit. It consist of three parts, a phase shifter, a variable amplifier and a controlled switch. The phase shifter delays the signal by $t=\pi/(2\Omega)$ effectively leading to a velocity feedback in the high-Q limit. The variable amplifier is used to change the strength of the feedback signal. The control switch is used to enable or disable the feedback control, which is used for the energy relaxation measurements.
		
		With appropriate settings of delay and gain the feedback laser applies a feedback force $F_\mathrm{fb}=m \Gamma_\mathrm{fb}\dot z$ where $\Gamma_\mathrm{fb}$ is proportional to the gain. The equation of motion becomes to
		\begin{align}
		\ddot z + (\Gamma_p+\Gamma_\mathrm{fb})\dot z + \Omega^2 z = \frac{F_\mathrm{therm}}{m}.
		\end{align}
		An effective temperature $T_\mathrm{COM}$ of the COM motion \cite{Poggio2007} can be defined 
		\begin{align}
		T_\mathrm{COM}	&=	T \frac{\Gamma_p}{\Gamma_p+\Gamma_\mathrm{fb}}
		\end{align}
		and for $\Gamma_p\ll\Gamma_\mathrm{fb}$ the expression simplifies to $T_\mathrm{COM}\approx T_0\Gamma_p/\Gamma_\mathrm{fb}$.
		
		\section{Transient behaviour of the potential energy}
		\label{ch:suppF}
\begin{figure}[h!]
			\centering
			\includegraphics{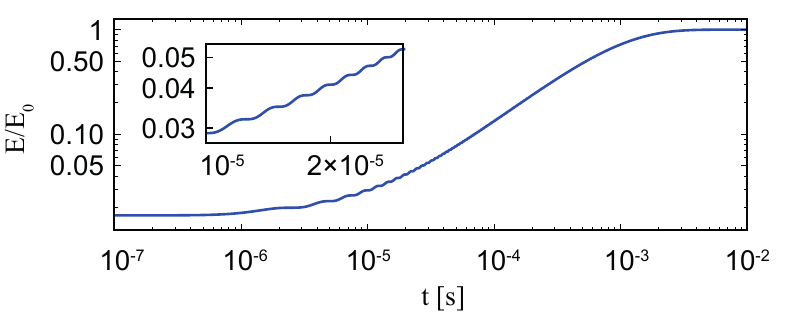}
			\caption{\label{fig:supp_energy}Thermalization of the potential energy from an initial state with $E=0.02E_0$ to thermal equilibrium}
		\end{figure}

		The equation of motion (equation 1, main text) describing the levitated nano-particle can be rewritten as a system of differential equations of first order introducing the velocity $v=\dot x$, a normalized state vector $\mathbf q=(x/x_\mathrm{th},v/v_\mathrm{th})^\top$ with $x_\mathrm{th}=\sqrt{2k_\mathrm{B} T/(m\Omega^2)}$ the RMS thermal amplitude, $v_\mathrm{th}=x_\mathrm{th}/\Omega$, $\mathbf n =(0,\sqrt{\Gamma_p})\zeta$ and $\zeta\sqrt{\Gamma_p}=F_\mathrm{therm}/(m\Omega x_\mathrm{th})$ to 
		\begin{align}
		\dot{\mathbf q}	&=	A\mathbf q + \mathbf n\zeta,
		\end{align}
		with A a constant matrix
		\begin{align}
		A	&=	\begin{pmatrix}
		0		&	\Omega \\
		-\Omega	&	-\Gamma_p
		\end{pmatrix}.
		\end{align}
		The time dependence of the covariance matrix $\Sigma(t)=\langle\mathbf q\mathbf q^\top\rangle$ is given by
		\begin{align}
		\dot\Sigma	&=	A\Sigma + \Sigma A^\top + \mathbf n\mathbf n^\top.
		\end{align}
		The first element of the covariance matrix contains the potential energy and is plotted in figure \ref{fig:supp_energy} for an initial state $E=0.02E_0$ (lowest temperature achieved, see main text) relaxing back to thermal equilibrium $E_0$.
		The energy relaxation from a cold state back to thermal equilibrium or from thermal equilibrium to a cold state in the main text is approximated with an exponential function. The analytical solution has an additional modulation, depicted in the inset in figure \ref{fig:supp_energy}, which is not resolved by our experimental data and not necessary to infer the relaxation constant $\Gamma_p$ or $\Gamma_\mathrm{fb}$.

		\section{Pressure distribution in the molecular flow regime}
		\label{ch:suppG}
		For a pressure reservoir at $p_1$ connected via a long round tube to a second pressure reservoir at $p_2<p_1$ the conductance $C$ of the tube in the molecular flow regime is $C=v\pi d^3/(12l)$ with $v$ the thermal gas velocity, $d$ the tube diameter and $l$ the tube length \cite{Fallis2013}. For a given pressure drop $\Delta p=p_1-p_2$ the flow $Q$ through the tube is $Q=C\Delta p$. As the conductance only depends on the pressure difference the pressure distribution inside a tube of length $l$ can be derived by considering a infinitesimal pressure drop $\delta p$ and length $\delta l$ 
		\begin{align}
		p(x)	&=	p_1-x\frac{12Q}{\pi v d^3}
		\end{align}
		with $x$ a position inside the tube $0<x<l$. Note that the dependence of the conductance on the pressure is different in different flow regimes, resulting in a modified pressure distribution \cite{Sharipov1998}.

	\nocite{*}
	\bibliography{library_x}

\end{document}